\documentclass[manuscript]{aastex}
\usepackage{amsmath}
\usepackage{natbib}
\shorttitle{K-shell photoionization of Na-like to Cl-like ions of Mg, Si, S, Ar,
            and Ca}              
\shortauthors{Witthoeft et~al.}

\begin{document}

\title{K-shell photoionization of Na-like to Cl-like ions of Mg, Si, S, Ar, 
       and Ca}
\author{M.C. Witthoeft, J. Garc\'ia, T.R. Kallman}
\affil{NASA Goddard Space Flight Center, Code 662, Greenbelt, MD 20771, USA}
\author{M.A. Bautista}
\affil{Department of Physics, Western Michigan University,\\
Kalamazoo, MI 49008, USA}
\author{C. Mendoza}
\affil{Centro de F\'{\i}sica, Instituto Venezolano de Investigaciones
Cient\'{\i}ficas (IVIC), \\ Caracas 1020A, Venezuela}
\author{P. Palmeri, P. Quinet}
\affil{Astrophysique et Spectroscopie, Universit\'{e} de Mons - UMONS, B-7000 
       Mons, Belgium}

%------------------------------------------------------------------------------

\begin{abstract} 
We present $R$-matrix calculations of photoabsorption and photoionization cross
sections across the K-edge of Mg, Si, S, Ar, and Ca ions with more than 10
electrons. 
The calculations include the effects of radiative and Auger damping by means
of an optical potential.
The wave functions are constructed from single-electron orbital bases obtained
using a Thomas--Fermi--Dirac statistical model potential. 
Configuration interaction is considered among all states up to $n=3$.
The damping processes affect the resonances converging to the K-thresholds
causing them to display symmetric profiles of constant width that smear the 
otherwise sharp edge at the photoionization threshold. 
These data are important for modeling of features found in photoionized plasmas.
\end{abstract}


\normalsize   

%------------------------------------------------------------------------------

\section{Introduction}

The current fleet of X-ray telescopes ({\em Chandra}, {\em XMM--Newton}, 
{\em Suzaku}) have revealed complex spectra from astrophysical sources such as
active galactic nuclei (AGNs) and X-ray binaries.
Besides the ubiquitous Fe absorption features, many other elements from C to Ni
have been observed in multiple ionization stages; see, for example, the 900
ksec Chandra spectrum of NGC 3783 from \citet{2002ApJ...574..643K}.
In order to accurately model systems producing such spectra, atomic data is
needed for all of these species, particularly the resonance structure leading
to the absorption features.
Due to the wavelengths of these features, the relevant atomic data involves the
K-shell of species from C to Ni where the currently available data has been of
limited accuracy.
Over the last several years, our group has worked on providing this data.
We have calculated energy levels, wavelengths, Einstein $A$-coefficients,
Auger rates, and photoionization cross sections.
Complete data sets have been computed for all ion stages of Fe 
\citep{2003A&A...403..339B,2003A&A...403.1175P, 2003A&A...410..359P,
2004A&A...414..377M,2004A&A...418.1171B, 2004ApJS..155..675K}, O 
\citep{2005ApJS..158...68G}, and N \citep{2009ApJS..185..477G}.
There has also been recent work on the C sequence by 
\citet{Hasogluetal.submitted.2010}.
This work is an extension of \citet{2008ApJS..177..408P} and 
\citet{2009ApJS..182..127W} where we present data for the Li-like to Ne-like 
ions of Ne, Mg, Si, S, Ar, and Ca.
Here we continue these calculations for the 23 non-neutral, Na-like to Cl-like 
ions of the same elements.

K-shell features of the ion stages covered here have been detected in X-ray
spectra of astronomical objects, most often due to fluorescent line emission.
Notable examples include the detection of K$\alpha$ emission from the ions 
Si {\sc II}-{\sc V} in the X-ray binary Vela X-1 \citep{2003ARA&A..41..291P}. 
This reveals the existence of a wide range of ionization states coexisting in
this system, and can best be explained if the emitting gas is highly clumped.
$K\alpha$ lines from S {\sc IV}-{\sc VI}, Ar {\sc VI}-{\sc VIII}, and Ca 
{\sc VI}-{\sc VII} have been detected in the X-ray binary 4U1700-37 
\citep{2003ApJ...592..516B}.  
The flux in these lines is correlated with the flux in the X-ray continuum, 
thus reinforcing the fluorescent origin.  
The K-edge of neutral Ne has been detected from interstellar gas along the line
of sight to bright X-ray binaries \citep{2003ApJ...599..498J} and provides
evidence for local enhancements in the interstellar Ne/O elemental abundance.
The data we present here is important both for its direct effect on observed
spectra via absorption features, and also through its effect on ionization
balance calculations, which in turn affect many of the observables from these 
elements.
In particular, it is the detailed resonance and edge structure, missing in
previous calculations, which is most important.

There have been a large number of calculations for these systems going back to
the 1920's, but they are confined to the non-resonant background and few cross
the K-edge region.
For the ions considered in this work, there is good agreement in the background
cross section both above and below the K-edge and are well characterized by the
fits of \citet{1995A&AS..109..125V}.
Measurements have been mainly confined to the neutral systems and for high
photon energies.
A comprehensive assessment and fitting of measurements has been performed by
\citet{1973AD......5...51V}.

There has been recent progress measuring the resonance structure near the 
K-edge of Li-like systems: C$^{3+}$ \citep{2009JPhB...42w5602M} and 
B$^{2+}$~\citep{2010JPhB...43m5602M}, as well as Be-like C$^{2+}$ 
\citep{2005JPhB...38.1967S}.
$R$-matrix calculations performed as part of those works are in good agreement
with the measurements apart from small discrepancies in some resonance
positions and heights.
With the advances of free electron lasers to produce X-rays
\citep{2006RaPC...75.2174K}, we expect it will be possible over the next
several years to perform similar experiments for the species under 
investigation here.

Tables of the total and partial cross sections accompany this article as on-line
tables. 
Full data sets are also available through the {\sc xstar} atomic database 
\citep{2001ApJS..134..139B}
\footnote{http://heasarc.nasa.gov/lheasoft/xstar/xstar.html}.

%------------------------------------------------------------------------------

\section{Numerical methods}

The $R$-matrix method employed here is the same as our previous K-shell
calculations and is described in detail in \citet{2003A&A...403..339B}.
The $R$-matrix method is based on the close-coupling approximation of
\citet{BurkeSeaton.MethCompPhys.1971} which is solved numerically following 
\citet{1971JPhB....4..153B,1974CoPhC...8..149B,1978CoPhC..14..367B,
1987JPhB...20.6379B}.
Due to the complexity of these M-shell ions, the scattering calculations are
carried out in $LS$-coupling except for the Na-like ions where we use the
Breit--Pauli $R$-matrix framework ({\sc bprm}).
The main limitation of the $LS$ calculations is that the results are not
fine-structure resolved.
As we will show in the next section, the total cross section is not sensitive
to the configuration (or coupling scheme) of the initial state.
However, for modeling purposes, we would like to have cross sections resolved
to the fine-structure level, so we split the cross section according to 
\citet{1976epia.conf..141R}.
Our wave functions were obtained using {\sc autostructure} 
\citep{1986JPhB...19.3827B,1997JPhB...30....1B} where the term (or level)
energies have been adjusted to match NIST \citep{NIST_Ralchenko} when possible.
For energies which are not available in NIST, most importantly those for the
K-hole states, we adjust to match the relativistic Hartree-Fock calculations of
\citet{2008ApJS..177..408P}.

As in our previous calculations, the effect of radiative and spectator Auger
decay on resonances is included using the optical potential of 
\citet{1996JPhB...29L.283G,2000JPhB...33.2511G}. 
Here, the resonance energy with respect to the threshold acquires an imaginary
component which is the sum of the radiative and Auger widths of the core. 
The radiative widths are computed within the $R$-matrix calculations following
\citet{1995PhRvA..52.1319R}. 
The Auger widths are provided from the relativistic Hartree-Fock calculations of
\citet{2008ApJS..177..408P}.
These effects are referred to as radiative and Auger damping since they cause
resonances in the photoionization cross section to become smaller in area.
The damping also makes it necessary to calculate both the photoabsorption and 
photoionization cross sections since not every absorbed photon will result in a
photoelectron.

We use the $R$-matrix computer package of \cite{1995CoPhC..92..290B} for the
inner region calculation and the asymptotic region code {\sc stgbf0damp} 
(Gorcyzca \& Badnell 1996; Badnell, unpublished) is used to compute the 
photoabsorption and photoionization cross sections including the effects of 
both radiation and Auger dampings. 
The target expansion for each ion includes the 1s$^2$ 2$l^8$ 3$l^N$,
1s$^2$ 2$l^7$ 3$l^{N+1}$, and 1s 2$l^8$ 3$l^{N+1}$ configurations where 
$N+10$ is the number of electrons in the photoionized (target) ion.
Only configurations with at most a single 3d electron are included.
The number of target states therefore can be as few as 43 levels (Na-like) or
as many as 132 terms (P-like).
We include enough continuum basis orbitals to span from threshold to at least
three times the energy of the K-edge.
This choice gives reliable cross sections up to at least 1.5 times the energy
of the K-edge.
Partial and total cross sections are calculated from all terms in the ground
configuration of the parent ion.

At this point, we would like to make a note about the accuracy of the K-edge
energy.  
For neutral systems where measurements exist, we can be fairly confident of the
edge position.
However, for ions, we must rely on calculated positions.
Due to the magnitude of these energies, differences on the order of a percent
yield uncertainties in the edge of a few Rydberg.
In the present calculations, the energy of the edge is determined by two 
quantities: the ionization potential calculated by $R$-matrix and the energy
of the first K-hole target state which we take from \citet{2008ApJS..177..408P}.
Since the ionization potential is known, we can (and do) adjust our cross
sections by the difference with observations ($\sim 0.2$ Ry) when transferring
our data to modeling codes.
It is more difficult to judge the accuracy of the K-vacancy target state 
energies except to say that we expect the accuracy to improve as the charge 
state increases; see \citet{2008ApJS..177..408P} for an assessment of the
accuracy.
We estimate that the resonance and edge positions of the singly-ionized systems
could be shifted by up to 1 Ry from their true values 
\citep{Gorczyca2010_private}.
We consider this to be the limiting error for these systems.
The accuracy of the edge and resonance positions should improve significantly
for the higher charge states.

\section{Results}

\begin{figure}
\includegraphics[width=10cm,angle=270]{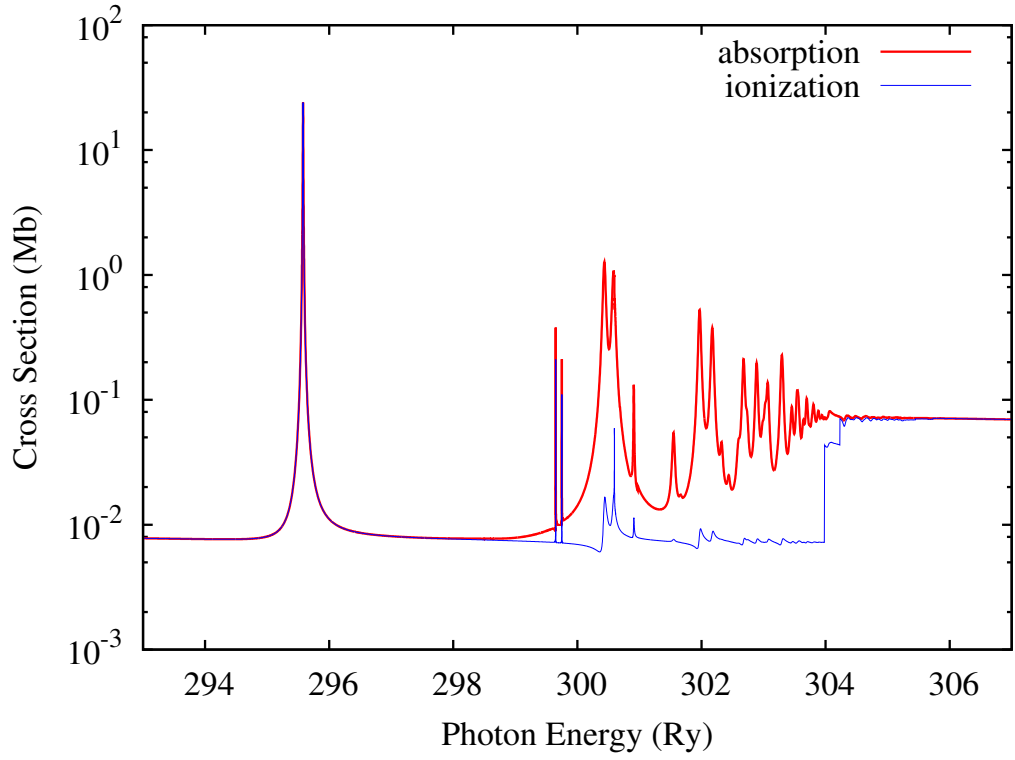}
\caption{
Photoabsorption (thick, red) and photoionization (thin, blue) cross sections
of P-like Ca near the K-edge.  Color available in the on-line version of the
text.
}
\label{fig:ca.plike}
\end{figure}

The cross sections reported here contain many of the same features shown in
calculations from other ions and elements.  
Inclusion of radiation and Auger damping broadens resonances near the K-edge
and makes them more symmetric.
The damping is also important in separating the absorption and ionization 
cross sections.
In Fig.~\ref{fig:ca.plike}, we show these cross sections for the $^4$S ground
term of P-like Ca.
There is little difference between absorption and ionization for the first, 
K$\beta$ resonance which indicates that damping plays little role here.
For the near-edge resonances, however, the photoionized resonances are nearly
damped away entirely while the strongly-broadened absorption resonances lead
to the characterstic smearing of the K-edge.
Small resonances can be observed above the edge in both absorption and 
ionization. 
These features are either typical resonances converging onto a higher-energy
edge or resonances belonging to doubly-excited states which are accessible
via mixing.
They are observed in all of the calculations, but rarely contribute much to 
the total cross section.

\begin{figure}
\includegraphics[width=10cm,angle=270]{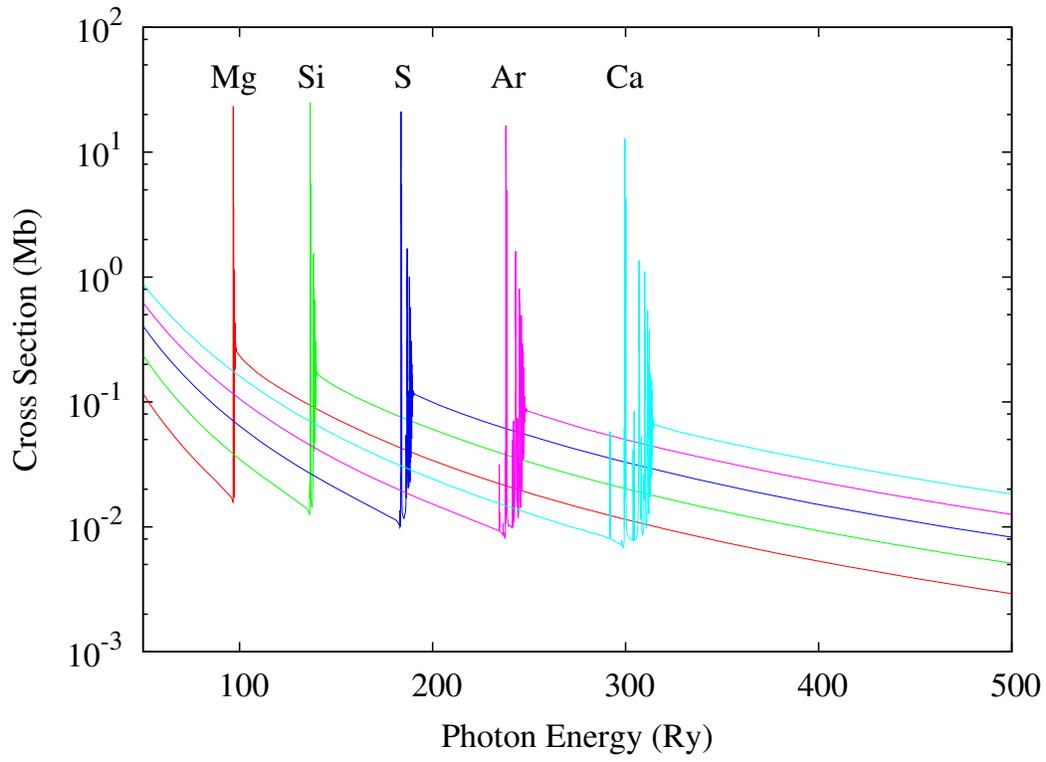}
\caption{
Photoabsorption cross sections for all Na-like ions included in this study.
Color available in the on-line version of the text.
}
\label{fig:isoelec.na}
\end{figure}

In Fig.~\ref{fig:isoelec.na}, we show the photoabsorption cross section for
each of the Na-like ions.
The behavior of the cross section across the iso-electronic sequence is quite
regular.
The background cross section above and below the K-edge increases with nuclear
charge and the width of the resonance region also increases with $Z$.
However the background cross section is known to be insensitive to the number
of electrons in the valence shell \citep{1977PhRvA..15.1001M,
1989PhRvA..40.6091N}.
If we were to plot the cross sections for all iso-electronic sequences on this
same figure, it would look much the same.
The exception is below the L-edge for sequences with more than 12 electrons
where the inner-shell (3s) electron has the same principal quantum number as
the valence (3p) electron.
In these cases, the insensitivity is lost \citep{2009PhRvA..80e3416P}.

\begin{figure}
\includegraphics[width=10cm,angle=270]{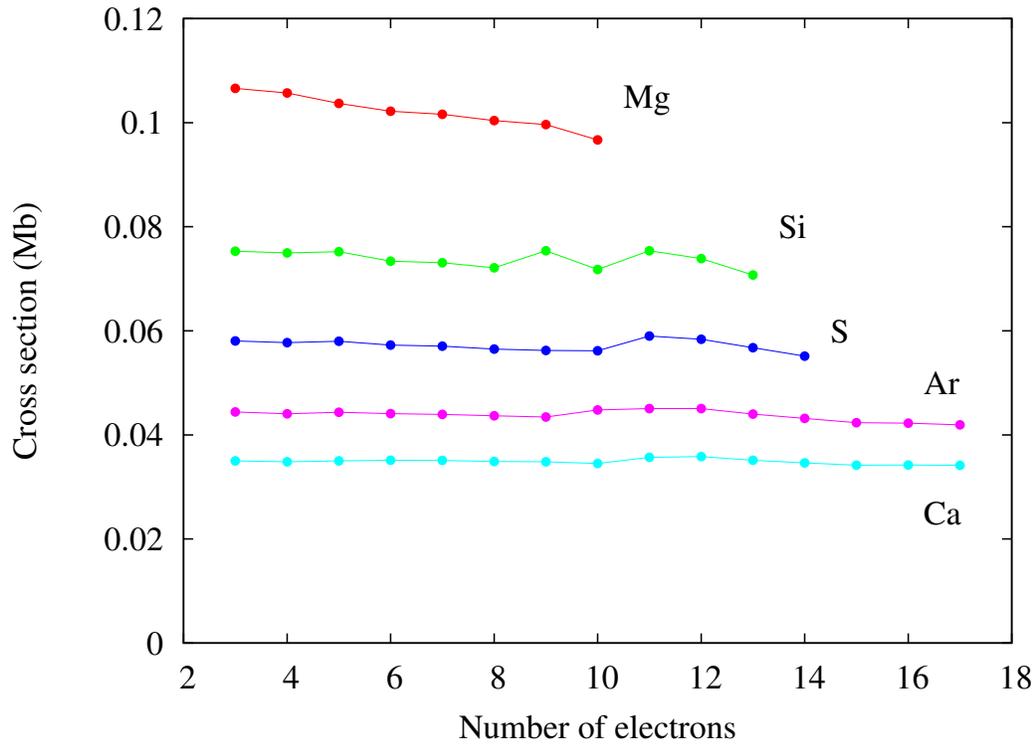}
\caption{
Background cross section at a fixed energy above the K-edge across iso-nuclear
sequences.  The photon energies for each sequence are: 128 Ry (Mg), 178 Ry (Si),
233 Ry (S), 300 Ry (Ar), and 375 Ry (Ca). Color available in the on-line
version of the text.
}
\label{fig:isonuclear}
\end{figure}

In Fig.~\ref{fig:isonuclear}, we show the background cross section at a fixed 
energy above the K-edge where results from \citet{2009ApJS..182..127W} are 
used for ions with 10 electrons or fewer.
We find the background cross section for each sequence to be constant within
10\%.
Note that the choice of energy is arbitrary and is different for each sequence
so no conclusions should be drawn about the relative cross sections between
sequences.
Nor does the choice of energy affect the agreement, we find better than 10\%
agreement between the cross sections at all energies where we can make
comparisons.
Only the Mg sequence shows a steady downward trend of the cross section with
an increasing number of electrons, but the total drop is only 10\% so it is 
difficult to make any conclusions.  
The remaining sequences show no trend with number of electrons.
Note that we are starting with the Li-like ions; if we also showed the cross
section for photoionization of the He-like ions, we would expect to see
differences since there is no inner-shell.

\begin{figure}
\includegraphics[width=10cm,angle=270]{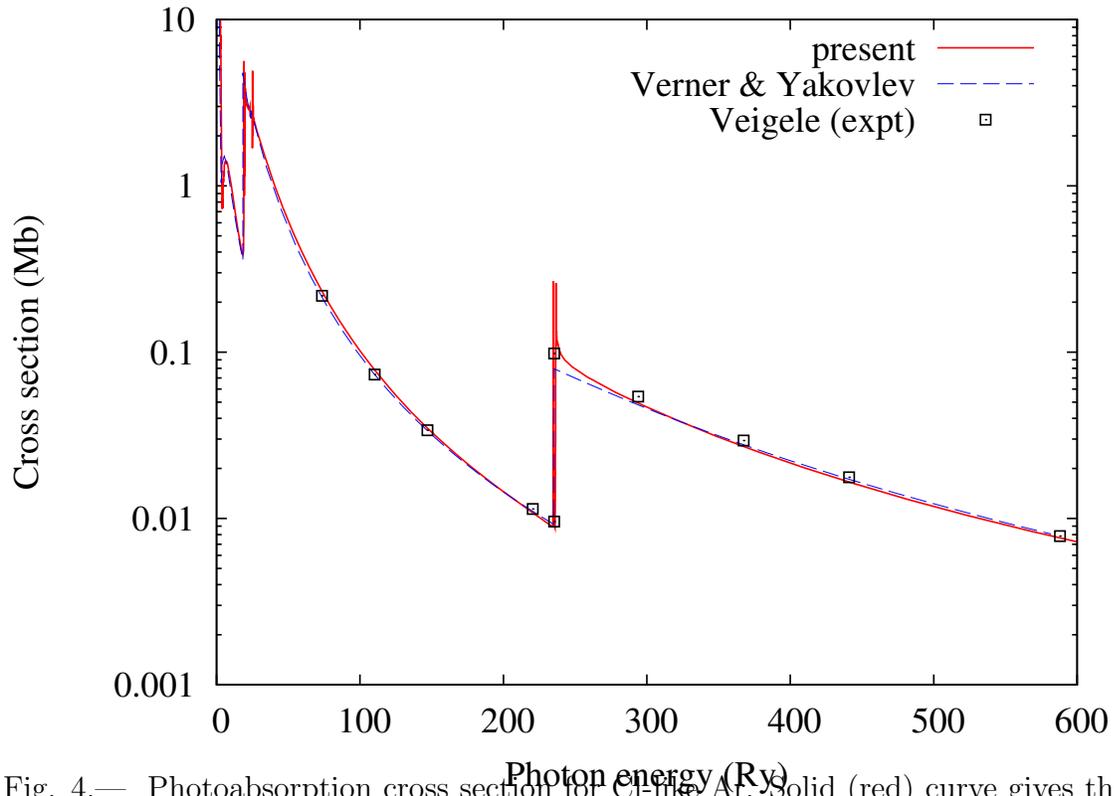}
\caption{
Photoabsorption cross section for Cl-like Ar.  Solid (red) curve gives the
present, $R$-matrix results, the dashed (blue) curve is from 
\citet{1995A&AS..109..125V}, and the boxes give the result from the 
experimental compilation of \citet{1973AD......5...51V}. Color available in 
the on-line version of the text.
}
\label{fig:ar.cllike}
\end{figure}

Previous experiments and calculations for K-shell photoionization were more
concerned with high-energy behavior than the detailed structure of the cross
section near the K-edge.
These results are generally in good agreement and have been characterized by
the fits of \citet{1995A&AS..109..125V}.
We want to compare our results with experiment, but measurements have only
been taken for neutral systems not calculated here.
However, we can take advantage of the finding above and compare our 
singly-ionized results with the neutral measurements as a test of the 
background.
In Fig.~\ref{fig:ar.cllike}, we show our cross section for Cl-like Ar compared
to the photoabsorption cross sections of neutral Ar given by 
\citet{1995A&AS..109..125V} and the fit to experimental data by 
\citet{1973AD......5...51V}.
There is very good agreement between the present results and 
\citet{1995A&AS..109..125V}
at all energies except at the K-edge where resonances give an enhancement to
the $R$-matrix results.
The experimental data point at the K-edge seems to confirm this enhancement,
although for the other sequences the Veigele data at the K-edge are in good
agreement with \citet{1995A&AS..109..125V}.
Except for the area in the immediate vicinity of the K-edge, there is 
similarly good agreement between all three results for the other elements.

Our results are to be used in the {\sc xstar} program 
\citep{2001ApJS..134..139B} for modeling photoionized plasmas.
To prepare for this application, we convolve our data with a Gaussian
and extrapolate to high energies.
As these calculations are focused only on the resonance features near the
K-edge, the cross section data from the Opacity Project 
\citep{1994MNRAS.266..805S} is used for the L-edge region.
The two data sets are joined together by hand at an energy between the L- and
K-edges.

The energy-dependent width used for the convolution is $\Delta E/E = 10^{-3}$
which is representative of current detectors.
The accuracy of $R$-matrix calculations starts to degrade at high energies
depending on the number of continuum basis functions included.
When this happens, we start to see oscillations in the cross section.
Therefore, the calculations are carried out to about 1.5 times the K-edge and
then extrapolated to 1100 Ry (or 15 keV).
The extrapolation assumes the asymptotic form, $\sigma(E) = A \, E^{-p}$ where
the parameters, $A$ and $p$, are determined by a least squares fit to the last
$\sim$100 points of the $R$-matrix data.
This method works well except for some weak partial cross section data where
we start to see oscillations.
To prevent unphysical extrapolation slopes, we enforce a lower limit of 8/3
on $p$ \citep{Cowan_PIasymp}.
While the extrapolation is not as accurate for these weak partials, it is not
likely to affect modeling.

\begin{figure}
\includegraphics[width=10cm,angle=270]{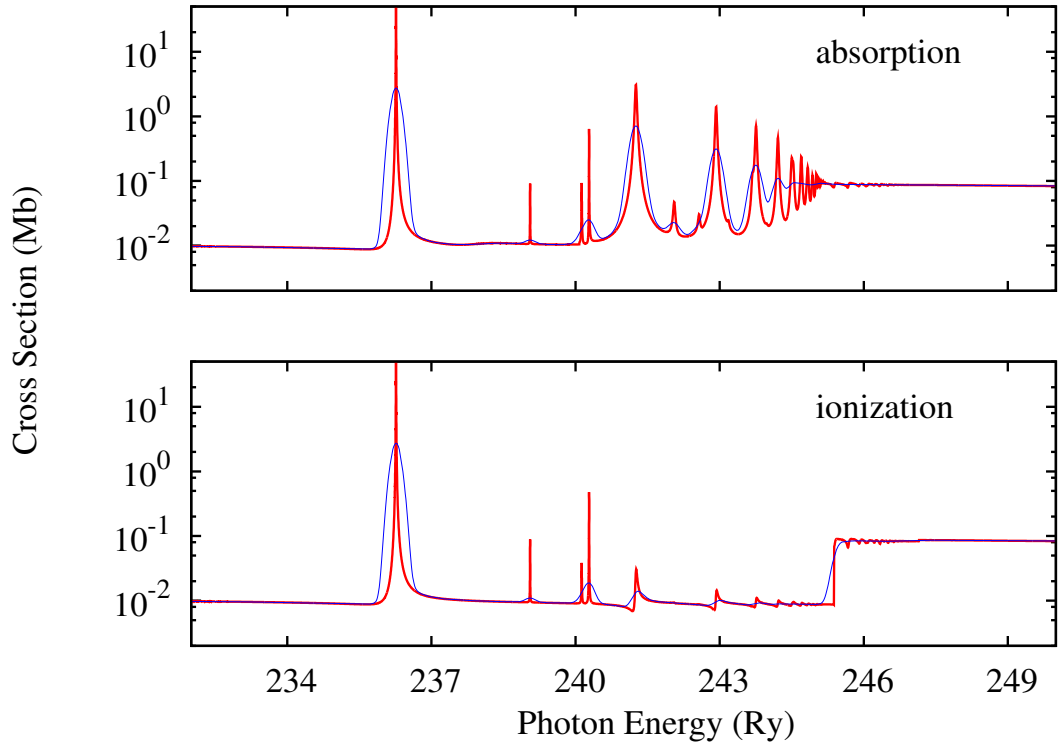}
\caption{
Raw (thick, red) and convolved (thin, blue) cross sections for Mg-like Ar near
the K-edge.
The top plot shows photoabsorption and the bottom is photoionization.
The convolution uses a width of $\Delta E/E = 0.001.$
Color available in the on-line version of the text.
}
\label{fig:convolve}
\end{figure}

Finally, after convolution and extrapolation, we remove unneccessary data points
from each data set.
Points which lie on a straight line between the two adjacent points within 1\%
are removed.
This test is repeated over the entire cross section until no more points are
removed.
Using this process, we can decrease the number of points in the cross section
data from several thousand to a couple hundred, yet still maintain good 
accuracy with linear interpolation.
In Fig.~\ref{fig:convolve}, we show the raw and convolved total cross sections
for Mg-like Ar near the K-edge.
The raw data contains over 11\,000 points while the convolved data has less
than 350 points.
The convolved cross sections are not included with this paper, but are
available by a request to MCW or {\sc xstar}.
The raw cross sections are available as electronic tables attached to this work.

%------------------------------------------------------------------------------

%-----------------------------------------------------------------------

\section{Summary and conclusions}

Total photoabsorption and photoionization cross sections have been computed 
for the K-shell of all non-neutral ions with 11 to 17 electrons for Mg, Si, S,
Ar, and Ca. 
Partial photoionization cross sections have also been calculated for the same
ions.
Radiative and spectator Auger dampings are accounted for in detail and the
energy region around the K-threshold was accurately treated for each ion.

We find good agreement in the background total cross section with experiment
\citep{1973AD......5...51V} and the fits of \citet{1995A&AS..109..125V}.
The data provides term-resolved partials which, to our knowledge, have not
previously been available.
For the Na-like ions, the present results are fully level-resolved.

All data are provided as on-line tables accompanying this article and can also
be obtained through the {\sc xstar} atomic database \citep{2001ApJS..134..139B} 
\footnotemark[\value{footnote}] and the Universal Atomic Database 
\footnote{http://heasarc.nasa.gov/uadb}.

The data sets provided here together with the energy levels and radiative and 
Auger rates reported in \citet{2008ApJS..177..408P} will help modelers to carry
out detail studies of K spectra of astrophysically abundant elements.

\begin{acknowledgments}
Support for this research was provided in part by a grant from the NASA 
Astronomy and Physics Research (APRA) program.
PP and PQ are respectively Research Associate and Senior Research Associate
of the Belgian F.R.S.-FNRS.  
Financial support from this organization is acknowledged.
\end{acknowledgments}

%----------------REFERENCES-----------------------------------------------------

\bibliography{mybib}

\begin{thebibliography}{41}
\expandafter\ifx\csname natexlab\endcsname\relax\def\natexlab#1{#1}\fi

\bibitem[{{Badnell}(1986)}]{1986JPhB...19.3827B}
{Badnell}, N.~R. 1986, Journal of Physics B Atomic Molecular Physics, 19, 3827

\bibitem[{{Badnell}(1997)}]{1997JPhB...30....1B}
---. 1997, Journal of Physics B Atomic Molecular Physics, 30, 1

\bibitem[{{Bautista} \& {Kallman}(2001)}]{2001ApJS..134..139B}
{Bautista}, M.~A., \& {Kallman}, T.~R. 2001, \apjs, 134, 139

\bibitem[{{Bautista} {et~al.}(2003){Bautista}, {Mendoza}, {Kallman}, \&
  {Palmeri}}]{2003A&A...403..339B}
{Bautista}, M.~A., {Mendoza}, C., {Kallman}, T.~R., \& {Palmeri}, P. 2003,
  \aap, 403, 339

\bibitem[{{Bautista} {et~al.}(2004){Bautista}, {Mendoza}, {Kallman}, \&
  {Palmeri}}]{2004A&A...418.1171B}
---. 2004, \aap, 418, 1171

\bibitem[{{Berrington} {et~al.}(1987){Berrington}, {Burke}, {Butler}, {Seaton},
  {Storey}, {Taylor}, \& {Yan}}]{1987JPhB...20.6379B}
{Berrington}, K.~A., {Burke}, P.~G., {Butler}, K., {Seaton}, M.~J., {Storey},
  P.~J., {Taylor}, K.~T., \& {Yan}, Y. 1987, Journal of Physics B Atomic
  Molecular Physics, 20, 6379

\bibitem[{{Berrington} {et~al.}(1974){Berrington}, {Burke}, {Chang}, {Chivers},
  {Robb}, \& {Taylor}}]{1974CoPhC...8..149B}
{Berrington}, K.~A., {Burke}, P.~G., {Chang}, J.~J., {Chivers}, A.~T., {Robb},
  W.~D., \& {Taylor}, K.~T. 1974, Computer Physics Communications, 8, 149

\bibitem[{{Berrington} {et~al.}(1978){Berrington}, {Burke}, {Le Dourneuf},
  {Robb}, {Taylor}, \& {Ky Lan}}]{1978CoPhC..14..367B}
{Berrington}, K.~A., {Burke}, P.~G., {Le Dourneuf}, M., {Robb}, W.~D.,
  {Taylor}, K.~T., \& {Ky Lan}, V. 1978, Computer Physics Communications, 14,
  367

\bibitem[{{Berrington} {et~al.}(1995){Berrington}, {Eissner}, \&
  {Norrington}}]{1995CoPhC..92..290B}
{Berrington}, K.~A., {Eissner}, W.~B., \& {Norrington}, P.~H. 1995, Computer
  Physics Communications, 92, 290

\bibitem[{{Boroson} {et~al.}(2003){Boroson}, {Vrtilek}, {Kallman}, \&
  {Corcoran}}]{2003ApJ...592..516B}
{Boroson}, B., {Vrtilek}, S.~D., {Kallman}, T., \& {Corcoran}, M. 2003, \apj,
  592, 516

\bibitem[{{Burke} {et~al.}(1971){Burke}, {Hibbert}, \&
  {Robb}}]{1971JPhB....4..153B}
{Burke}, P.~G., {Hibbert}, A., \& {Robb}, W.~D. 1971, Journal of Physics B
  Atomic Molecular Physics, 4, 153

\bibitem[{{Burke} \& {Seaton}(1971)}]{BurkeSeaton.MethCompPhys.1971}
{Burke}, P.~G., \& {Seaton}, M.~J. 1971, Methods in Computational Physics,
  Vol.~10, Numerical Solution of the Integro-Differential Equations of
  Electron-Atom Collision Theory (Academic Press), 1

\bibitem[{{Cowan}(1981)}]{Cowan_PIasymp}
{Cowan}, R.~D. 1981, The Theory of Atomic Structure and Spectra (University of
  California Press), 526

\bibitem[{{Garc{\'{\i}}a} {et~al.}(2005){Garc{\'{\i}}a}, {Mendoza}, {Bautista},
  {Gorczyca}, {Kallman}, \& {Palmeri}}]{2005ApJS..158...68G}
{Garc{\'{\i}}a}, J., {Mendoza}, C., {Bautista}, M.~A., {Gorczyca}, T.~W.,
  {Kallman}, T.~R., \& {Palmeri}, P. 2005, \apjs, 158, 68

\bibitem[{{Garc{\'{\i}}a} {et~al.}(2009){Garc{\'{\i}}a}, {Kallman},
  {Witthoeft}, {Behar}, {Mendoza}, {Palmeri}, {Quinet}, {Bautista}, \&
  {Klapisch}}]{2009ApJS..185..477G}
{Garc{\'{\i}}a}, J., {et~al.} 2009, \apjs, 185, 477

\bibitem[{{Gorczyca}(2010)}]{Gorczyca2010_private}
{Gorczyca}, T.~W. 2010, private communication

\bibitem[{{Gorczyca} \& {Badnell}(1996)}]{1996JPhB...29L.283G}
{Gorczyca}, T.~W., \& {Badnell}, N.~R. 1996, Journal of Physics B Atomic
  Molecular Physics, 29, L283

\bibitem[{{Gorczyca} \& {Badnell}(2000)}]{2000JPhB...33.2511G}
---. 2000, Journal of Physics B Atomic Molecular Physics, 33, 2511

\bibitem[{{Hasoglu} {et~al.}(2010){Hasoglu}, {Abdel-Naby}, {Gorczyca}, {Drake},
  \& {McLaughlin}}]{Hasogluetal.submitted.2010}
{Hasoglu}, M.~F., {Abdel-Naby}, S.~A., {Gorczyca}, T.~W., {Drake}, J.~J., \&
  {McLaughlin}, B.~M. 2010, \apj, submitted, available at arXiv:1003.3639v1

\bibitem[{{Juett} \& {Chakrabarty}(2003)}]{2003ApJ...599..498J}
{Juett}, A.~M., \& {Chakrabarty}, D. 2003, \apj, 599, 498

\bibitem[{{Kallman} {et~al.}(2004){Kallman}, {Palmeri}, {Bautista}, {Mendoza},
  \& {Krolik}}]{2004ApJS..155..675K}
{Kallman}, T.~R., {Palmeri}, P., {Bautista}, M.~A., {Mendoza}, C., \& {Krolik},
  J.~H. 2004, \apjs, 155, 675

\bibitem[{{Kanter} {et~al.}(2006){Kanter}, {Dunford}, {Kr{\"a}ssig},
  {Southworth}, \& {Young}}]{2006RaPC...75.2174K}
{Kanter}, E.~P., {Dunford}, R.~W., {Kr{\"a}ssig}, B., {Southworth}, S.~H., \&
  {Young}, L. 2006, Radiation Physics and Chemistry, 75, 2174

\bibitem[{{Kaspi} {et~al.}(2002){Kaspi}, {Brandt}, {George}, {Netzer},
  {Crenshaw}, {Gabel}, {Hamann}, {Kaiser}, {Koratkar}, {Kraemer}, {Kriss},
  {Mathur}, {Mushotzky}, {Nandra}, {Peterson}, {Shields}, {Turner}, \&
  {Zheng}}]{2002ApJ...574..643K}
{Kaspi}, S., {et~al.} 2002, \apj, 574, 643

\bibitem[{{Mendoza} {et~al.}(2004){Mendoza}, {Kallman}, {Bautista}, \&
  {Palmeri}}]{2004A&A...414..377M}
{Mendoza}, C., {Kallman}, T.~R., {Bautista}, M.~A., \& {Palmeri}, P. 2004,
  \aap, 414, 377

\bibitem[{{Missavage} {et~al.}(1977){Missavage}, {Manson}, \&
  {Daum}}]{1977PhRvA..15.1001M}
{Missavage}, D.~W., {Manson}, S.~T., \& {Daum}, G.~R. 1977, \pra, 15, 1001

\bibitem[{{M{\"u}ller} {et~al.}(2009){M{\"u}ller}, {Schippers}, {Phaneuf},
  {Scully}, {Aguilar}, {Covington}, {{\'A}lvarez}, {Cisneros}, {Emmons},
  {Gharaibeh}, {Hinojosa}, {Schlachter}, \& {McLaughlin}}]{2009JPhB...42w5602M}
{M{\"u}ller}, A., {et~al.} 2009, Journal of Physics B Atomic Molecular Physics,
  42, 235602

\bibitem[{{M{\"u}ller} {et~al.}(2010){M{\"u}ller}, {Schippers}, {Phaneuf},
  {Scully}, {Aguilar}, {Cisneros}, {Gharaibeh}, {Schlachter}, \&
  {McLaughlin}}]{2010JPhB...43m5602M}
---. 2010, Journal of Physics B Atomic Molecular Physics, 43, 135602

\bibitem[{{Nasreen} {et~al.}(1989){Nasreen}, {Manson}, \&
  {Deshmukh}}]{1989PhRvA..40.6091N}
{Nasreen}, G., {Manson}, S.~T., \& {Deshmukh}, P.~C. 1989, \pra, 40, 6091

\bibitem[{{Paerels} \& {Kahn}(2003)}]{2003ARA&A..41..291P}
{Paerels}, F.~B.~S., \& {Kahn}, S.~M. 2003, \araa, 41, 291

\bibitem[{{Palmeri} {et~al.}(2003{\natexlab{a}}){Palmeri}, {Mendoza},
  {Kallman}, \& {Bautista}}]{2003A&A...403.1175P}
{Palmeri}, P., {Mendoza}, C., {Kallman}, T.~R., \& {Bautista}, M.~A.
  2003{\natexlab{a}}, \aap, 403, 1175

\bibitem[{{Palmeri} {et~al.}(2003{\natexlab{b}}){Palmeri}, {Mendoza},
  {Kallman}, {Bautista}, \& {Mel{\'e}ndez}}]{2003A&A...410..359P}
{Palmeri}, P., {Mendoza}, C., {Kallman}, T.~R., {Bautista}, M.~A., \&
  {Mel{\'e}ndez}, M. 2003{\natexlab{b}}, \aap, 410, 359

\bibitem[{{Palmeri} {et~al.}(2008){Palmeri}, {Quinet}, {Mendoza}, {Bautista},
  {Garc{\'{\i}}a}, \& {Kallman}}]{2008ApJS..177..408P}
{Palmeri}, P., {Quinet}, P., {Mendoza}, C., {Bautista}, M.~A., {Garc{\'{\i}}a},
  J., \& {Kallman}, T.~R. 2008, \apjs, 177, 408

\bibitem[{{Pradhan} {et~al.}(2009){Pradhan}, {Jose}, {Deshmukh},
  {Radojevi{\'c}}, \& {Manson}}]{2009PhRvA..80e3416P}
{Pradhan}, G.~B., {Jose}, J., {Deshmukh}, P.~C., {Radojevi{\'c}}, V., \&
  {Manson}, S.~T. 2009, \pra, 80, 053416

\bibitem[{{Ralchenko} {et~al.}(2008){Ralchenko}, {Kramida}, {Reader}, \& {NIST
  ASD Team}}]{NIST_Ralchenko}
{Ralchenko}, Y., {Kramida}, A.~E., {Reader}, J., \& {NIST ASD Team}. 2008, NIST
  Atomic Spectra Database (version 4.0) (National Institute of Standards and
  Technology, Gaithersburg, MD), http://physics.nist.gov/asd3

\bibitem[{{Rau}(1976)}]{1976epia.conf..141R}
{Rau}, A.~R.~P. 1976, in Electron and Photon Interactions with Atoms, ed.
  {H.~Kleinpoppen \& M.~R.~C.~McDowell}, 141--+

\bibitem[{{Robicheaux} {et~al.}(1995){Robicheaux}, {Gorczyca}, {Pindzola}, \&
  {Badnell}}]{1995PhRvA..52.1319R}
{Robicheaux}, F., {Gorczyca}, T.~W., {Pindzola}, M.~S., \& {Badnell}, N.~R.
  1995, \pra, 52, 1319

\bibitem[{{Scully} {et~al.}(2005){Scully}, {Aguilar}, {Emmons}, {Phaneuf},
  {Halka}, {Leitner}, {Levin}, {Lubell}, {P{\"u}ttner}, {Schlachter},
  {Covington}, {Schippers}, {M{\"u}ller}, \&
  {McLaughlin}}]{2005JPhB...38.1967S}
{Scully}, S.~W.~J., {et~al.} 2005, Journal of Physics B Atomic Molecular
  Physics, 38, 1967

\bibitem[{{Seaton} {et~al.}(1994){Seaton}, {Yan}, {Mihalas}, \&
  {Pradhan}}]{1994MNRAS.266..805S}
{Seaton}, M.~J., {Yan}, Y., {Mihalas}, D., \& {Pradhan}, A.~K. 1994, \mnras,
  266, 805

\bibitem[{{Veigele}(1973)}]{1973AD......5...51V}
{Veigele}, W.~J. 1973, Atomic Data, 5, 51

\bibitem[{{Verner} \& {Yakovlev}(1995)}]{1995A&AS..109..125V}
{Verner}, D.~A., \& {Yakovlev}, D.~G. 1995, \aaps, 109, 125

\bibitem[{{Witthoeft} {et~al.}(2009){Witthoeft}, {Bautista}, {Mendoza},
  {Kallman}, {Palmeri}, \& {Quinet}}]{2009ApJS..182..127W}
{Witthoeft}, M.~C., {Bautista}, M.~A., {Mendoza}, C., {Kallman}, T.~R.,
  {Palmeri}, P., \& {Quinet}, P. 2009, \apjs, 182, 127

\end{thebibliography}

%---------------------Figs -------------------------------------------------

%-------------------------------------------------------------------------------

\end{document}